\documentclass[english,aps,superscriptaddress,preprint]{revtex4}
\usepackage[T1]{fontenc}
\usepackage{color}
\usepackage{array}
\usepackage{multirow}
\usepackage{amsmath}
\usepackage{amssymb}
\usepackage{graphicx}

\makeatletter

\providecommand{\tabularnewline}{\\}

\@ifundefined{textcolor}{}
{%
 \definecolor{BLACK}{gray}{0}
 \definecolor{WHITE}{gray}{1}
 \definecolor{RED}{rgb}{1,0,0}
 \definecolor{GREEN}{rgb}{0,1,0}
 \definecolor{BLUE}{rgb}{0,0,1}
 \definecolor{CYAN}{cmyk}{1,0,0,0}
 \definecolor{MAGENTA}{cmyk}{0,1,0,0}
 \definecolor{YELLOW}{cmyk}{0,0,1,0}
}

\usepackage{dcolumn}
\usepackage{bm}
\usepackage{times}
\usepackage{hyperref}

\makeatother

\usepackage{babel}
\begin{document}
\title{Electron parallel closures for various ion charge numbers\\ \href{https://dx.doi.org/10.1063/1.4944665}{\small Journal-ref:  \underline{Phys.  Plasmas 23, 032124 (2016)}} {\small \color{red}  with corrections} }
\author{Jeong-Young Ji}
\email{j.ji@usu.edu}

\address{Department of Physics, Utah State University, Logan, Utah 84322}
\author{Sang-Kyeun Kim}
\address{Department of Nuclear Engineering, Seoul National University, Seoul
151-742, Korea}
\author{Eric D. Held}
\address{Department of Physics, Utah State University, Logan, Utah 84322}
\author{Yong-Su Na}
\address{Department of Nuclear Engineering, Seoul National University, Seoul
151-742, Korea}
\begin{abstract}
Electron parallel closures for the ion charge number $Z=1$ {[}J.-Y.
Ji and E. D. Held, Phys. Plasmas \textbf{21}, 122116 (2014){]} are
extended for $1\le Z\le10$. Parameters are computed for various $Z$
with the same form of the $Z=1$ kernels adopted. The parameters are
smoothly varying in $Z$ and hence can be used to interpolate parameters
and closures for noninteger, effective ion charge numbers. 
\end{abstract}
\maketitle

\section{Introduction}

A set of fluid equations for density $(n)$, temperature ($T$), and
flow velocity ($\mathbf{V}$) require closure relations for heat flux
density (\textbf{$\mathbf{h}$}), friction force density ($\mathbf{R}$),
and viscous pressure tensor $(\boldsymbol{\pi})$. For electron-ion
plasmas in a magnetic field, a complete set of closures has been obtained
for high collisionality~\citep{Braginskii1958,Braginskii1965}. In
a magnetized plasma, parallel closures for moderate- and low-collisionality
plasma are studied with approximate collision operators in Refs.~\citep{Chang1992C,Held2001CHS,Held2003,Held2003CH,Held2004-5}.
Accurate collision operators~\citep{Ji2009HS,Ji2009H1} are adopted
in the general moment approach~\citep{Ji2009HS,Ji2009H1}. In general,
the parallel closures are expressed by kernel-weighted integrals.
The kernels obtained from the moment method appear in a series of
exponential functions and are valid up to moderately low collisionality
depending on the number of moments. Closures in the collisionless
limit have been studied in Refs.~\citep{Hammett1990P,Chang1992C,Hazeltine1998,Ji2013HJ}. 

From the moment kernels and collisionless kernels, simple fitted kernels
for arbitrary collisionality are obtained for $Z=1$ in Ref.~\citep{Ji2014H1}.
For completeness and application to various ion charge numbers~\citep{Lee2000e,Rathgeber2010e,Kallenbach2013e},
we extend the $Z=1$ work to $1<Z\le10$. The fitted kernels are specified
by seven parameters and the parameters have many local minima in the
least square fitting. Among them we choose minima where parameters
change smoothly in $Z$. The smoothness enables us to compute kernel
parameters and closures for a noninteger effective ion charge number
$Z_{\mathrm{eff}}$. 

In Sec.~\ref{sec:MEll}, we review the parallel moment equations
and the properties of kernels for the integral closures. In Sec.~\ref{sec:pll},
the fitted kernel parameters and accuracy of closures are presented
for $1\le Z\le10$. In Sec.~\ref{sec:Dis} we summarize.

\section{parallel moment equations and integral closures\label{sec:MEll}}

To obtain closures for the Maxwellian (M) moment equations, we decompose
a distribution function into the Maxwellian ($f^{\mathrm{M}})$ and
non-Maxwellian parts $(f^{\mathrm{N}})$, and then solve a reduced
(approximate) kinetic equation for $f^{\mathrm{N}}$. For \emph{parallel}
closures, we solve a drift kinetic equation to find a gyro-averaged
distribution function $(\bar{f})$,
\begin{equation}
v_{\|}\frac{\partial\bar{f}_{\mathrm{e}}^{\mathrm{N}}}{\partial\ell}=\overline{C_{\mathrm{eL}}(f_{\mathrm{e}}^{\mathrm{N}})}-v_{\|}\frac{\partial\bar{f}_{\mathrm{e}}^{\mathrm{M}}}{\partial\ell}+\overline{C_{\mathrm{eL}}(f_{\mathrm{e}}^{\mathrm{M}})}\label{rKE}
\end{equation}
for $\bar{f}_{\mathrm{e}}^{\mathrm{N}}$ in terms of $\bar{f}_{\mathrm{e}}^{\mathrm{M}}$,
where $C_{\mathrm{eL}}$ is the linearized Landau-Fokker-Planck operator
with respect to $f_{a=\mathrm{e},\mathrm{i}}^{\mathrm{M}}$ for an
electron distribution function $F$,
\begin{equation}
C_{\mathrm{eL}}(F)=C(F,f_{\mathrm{e}}^{\mathrm{M}})+C(f_{\mathrm{e}}^{\mathrm{M}},F)+C(F,f_{\mathrm{i}}^{\mathrm{M}}).\label{CeL}
\end{equation}
When solving Eq.~\eqref{rKE} for closures, we must remove the fluid
moment equations to be closed~\citep{Ji2013HJ}. 

In the total-velocity moment expansion, the distribution functions
are 
\begin{equation}
f_{a}^{\mathrm{M}}\approx f_{a}^{\mathrm{m}}(1+2\mathbf{s}_{a}\cdot\frac{\mathbf{V}_{a}}{v_{Ta}})=f_{a}^{\mathrm{m}}+f_{a}^{\mathrm{M}-\mathrm{m}},
\end{equation}
\begin{equation}
f_{a}^{\mathrm{N}}=f_{a}^{\mathrm{m}}\sum_{lk\ne\mathrm{M}}\hat{\mathsf{P}}_{a}^{lk}\cdot\mathsf{M}_{a}^{lk},
\end{equation}
with 
\begin{equation}
f_{a}^{\mathrm{m}}=n_{a}\hat{f}_{a}^{\mathrm{m}},\;\hat{f}_{a}^{\mathrm{m}}=\frac{1}{\pi^{3/2}v_{Ta}^{3}}e^{-s_{a}^{2}},
\end{equation}
and
\begin{equation}
\hat{\mathsf{P}}_{a}^{lk}=\frac{1}{\sqrt{\sigma_{l}\lambda_{k}^{l}}}\mathsf{P}_{a}^{lk},\;\mathsf{P}_{a}^{lk}=\mathsf{P}^{l}(\mathbf{s}_{a})L_{k}^{(l+1/2)}(s_{a}^{2}).
\end{equation}
Here $\mathbf{s}_{a}=\mathbf{v}/v_{Ta}$, $v_{Ta}=\sqrt{2T_{a}/m_{a}}$,
$\mathbf{V}_{a}$ is the flow velocity, $\sigma_{l}=l!/(2l+1)!!$,
$\lambda_{k}^{l}=(l+k+1/2)!/k!(1/2)!$, $\mathsf{P}^{l}$ is a harmonic
tensor, and $L_{k}^{(l+1/2)}$ is a Laguerre-Sonine polynomial. Now
the collision operators can be further linearized with respect to
$f_{\mathrm{e}}^{\mathrm{m}}$ and $f_{\mathrm{i}}^{\mathrm{m}}$,
\begin{equation}
C_{\mathrm{eL}}(f_{\mathrm{e}}^{\mathrm{N}})\approx C(f_{\mathrm{e}}^{\mathrm{N}},f_{\mathrm{e}}^{\mathrm{m}})+C(f_{\mathrm{e}}^{\mathrm{m}},f_{\mathrm{e}}^{\mathrm{N}})+C(f_{\mathrm{e}}^{\mathrm{N}},f_{\mathrm{i}}^{\mathrm{m}}),\label{CeLN}
\end{equation}
 and 
\begin{equation}
C_{\mathrm{eL}}(f_{\mathrm{e}}^{\mathrm{M}})=C(f_{\mathrm{e}}^{\mathrm{M}},f_{\mathrm{i}}^{\mathrm{M}})\approx C(f_{\mathrm{e}}^{\mathrm{m}},f_{\mathrm{i}}^{\mathrm{m}})+C(f_{\mathrm{e}}^{\mathrm{M}-\mathrm{m}},f_{\mathrm{i}}^{\mathrm{m}})+C(f_{\mathrm{e}}^{\mathrm{m}},f_{\mathrm{i}}^{\mathrm{M}-\mathrm{m}}).\label{CeLM}
\end{equation}

The gyro-averaged distribution function, $\bar{f}=(2\pi)^{-1}\int d\gamma f$
where $\gamma$ is the gyro-angle, can be written as
\begin{equation}
\bar{f}_{a}^{\mathrm{M}}\approx f_{a}^{\mathrm{m}}(1+2s_{a\|}\frac{V_{a\|}}{v_{Ta}})=f_{a}^{\mathrm{m}}+\bar{f}_{\mathrm{e}}^{\mathrm{M}-\mathrm{m}},
\end{equation}
\begin{equation}
\bar{f}_{a}^{\mathrm{N}}=f_{a}^{\mathrm{m}}\sum_{lk\ne\mathrm{M}}\hat{P}_{a}^{lk}n_{a}^{lk},
\end{equation}
with 
\begin{equation}
\hat{P}_{a}^{lk}=\frac{1}{\sqrt{\bar{\sigma}_{l}\lambda{}_{k}^{l}}}P_{a}^{lk},\;P_{a}^{lk}=s_{a}^{l}P_{l}(\xi)L_{k}^{(l+1/2)}(s_{a}^{2}),
\end{equation}
\begin{equation}
n_{a}^{lk}=\sqrt{\frac{\bar{\sigma}_{l}}{\sigma_{l}}}n_{a}M_{a\|}^{lk},
\end{equation}
where $\bar{\sigma}_{l}=1/(2l+1)$, $\xi=v_{\|}/v$, and $P_{l}$
is a Legendre polynomial. It has been shown~\citep{Ji2006H} that
the gyroaverage of the linearized operators of distribution functions,
Eqs.~\eqref{CeLN} and \eqref{CeLM}, are the same as the linearized
operators of the gyroaveraged distribution functions, i.e., 
\begin{equation}
\overline{C_{\mathrm{eL}}(f_{\mathrm{e}}^{\mathrm{N}})}\approx C(\bar{f}_{\mathrm{e}}^{\mathrm{N}},f_{\mathrm{e}}^{\mathrm{m}})+C(f_{\mathrm{e}}^{\mathrm{m}},\bar{f}_{\mathrm{e}}^{\mathrm{N}})+C(\bar{f}_{\mathrm{e}}^{\mathrm{N}},f_{\mathrm{i}}^{\mathrm{m}})\label{CeLNa}
\end{equation}
 and 
\begin{equation}
\overline{C_{\mathrm{eL}}(f_{\mathrm{e}}^{\mathrm{M}})}\approx C(f_{\mathrm{e}}^{\mathrm{m}},f_{\mathrm{i}}^{\mathrm{m}})+C(\bar{f}_{\mathrm{e}}^{\mathrm{M}-\mathrm{m}},f_{\mathrm{i}}^{\mathrm{m}})+C(f_{\mathrm{e}}^{\mathrm{m}},\bar{f}_{\mathrm{i}}^{\mathrm{M}-\mathrm{m}}).\label{CeLMa}
\end{equation}

To obtain the $(j,p)$ moment equation, we multiply $\hat{P}^{jp}$
to Eq.~\eqref{rKE} and integrate over velocity space
\begin{equation}
v_{T}\sum_{lk\ne\mathrm{M}}\bar{\Psi}^{jp,lk}\frac{\partial n^{lk}}{\partial\ell}=\frac{1}{\tau_{\mathrm{ee}}}\sum_{lk\ne\mathrm{M}}c^{jp,lk}n^{lk}+\frac{1}{\tau_{\mathrm{ee}}}g^{jp},\label{ME2}
\end{equation}
where 
\begin{equation}
\bar{\Psi}^{jp,lk}=\int d\mathbf{v}\hat{P}^{jp}s_{\|}\hat{f}^{\mathrm{m}}\hat{P}^{lk},\label{Psi}
\end{equation}
\begin{equation}
c^{jp,lk}=\tau_{\mathrm{ee}}\int d\mathbf{v}\hat{P}^{jp}C_{\mathrm{eL}}(\hat{f}^{\mathrm{m}}\hat{P}^{lk})=\delta_{jl}c_{pk}^{j},\label{c}
\end{equation}
and 
\begin{equation}
g^{jp}=\int d\mathbf{v}\hat{P}^{jp}[\tau_{\mathrm{ee}}\overline{C_{\mathrm{eL}}(f_{\mathrm{e}}^{\mathrm{M}})}-\lambda_{\mathrm{C}}s_{\|}\frac{\partial\bar{f}_{\mathrm{e}}^{\mathrm{M}}}{\partial\ell}],\label{g}
\end{equation}
where $\lambda_{\mathrm{C}}=v_{T}\tau_{\mathrm{ee}}$ and $\tau_{\mathrm{ee}}$
is the electron-electron collision time. The electron collision matrix
can be computed from 
\begin{equation}
c_{pk}^{j}=\frac{\tau_{\mathrm{e}\mathrm{e}}}{n_{\mathrm{e}}\sqrt{\lambda_{p}^{j}\lambda_{k}^{j}}}(A_{\mathrm{ee}}^{jpk}+B_{\mathrm{ee}}^{jpk}+A_{\mathrm{ei}}^{jpk})
\end{equation}
where 
\begin{eqnarray}
\bar{\sigma}_{j}A_{ab}^{jpk} & = & \int d\mathbf{v}P_{a}^{jp}C(f_{a}^{\mathrm{m}}P_{a}^{jk},f_{b}^{\mathrm{m}}),\\
\bar{\sigma}_{j}B_{ab}^{jpk} & = & \int d\mathbf{v}P_{a}^{jp}C(f_{a}^{\mathrm{m}},f_{b}^{\mathrm{m}}P_{a}^{jk}).
\end{eqnarray}
and formulae for $A_{ab}^{jpk}$ and $B_{ab}^{jpk}$ are presented
in Refs.~\citep{Ji2006H,Ji2008H}. For electrons, the nonvanishing
thermodynamic drives $g_{A}$ are 
\begin{eqnarray}
g^{1k} & = & \delta_{1k}\frac{\sqrt{5}}{2}\frac{n}{T}\frac{dT}{d\eta}+\sqrt{2}Za_{\mathrm{ei}}^{1k0}n\hat{V}_{\mathrm{ei}\|},\label{g1k}\\
g^{20} & = & -\frac{\sqrt{3}}{2}n\tau_{\mathrm{ee}}W_{\|},\label{g20}
\end{eqnarray}
where
\begin{equation}
a_{\mathrm{ei}}^{10k}=a_{\mathrm{ei}}^{1k0}=-\sqrt{\frac{3(k+1/2)!}{(2k+3)k!(1/2)!}},
\end{equation}
\begin{equation}
\hat{V}_{\mathrm{ei}\|}=\frac{\mathbf{b}\cdot(\mathbf{V}_{\mathrm{e}}-\mathbf{V}_{\mathrm{i}})}{v_{T}},
\end{equation}
and
\begin{equation}
W_{\|}=\mathbf{b}\mathbf{b}:\mathsf{W},\;(\mathsf{W})_{\alpha\beta}=\partial_{\alpha}V_{\beta}+\partial_{\beta}V_{\alpha}-\frac{2}{3}\delta_{\alpha\beta}\nabla\cdot\mathbf{V}.
\end{equation}
The parallel closures are related to the general moments by 
\begin{eqnarray}
h_{\|} & = & -\frac{\sqrt{5}}{2}v_{T}Tn^{11},\label{h:n}\\
R_{\|} & = & \frac{m_{\mathrm{e}}v_{T\mathrm{e}}}{\tau_{\mathrm{ei}}}[-n_{\mathrm{e}}\hat{V}_{\mathrm{ei}\|}+\frac{1}{\sqrt{2}}\sum_{k=1}a_{\mathrm{ei}}^{10k}n^{1k}],\label{R:n}\\
\pi_{\|} & = & \frac{2}{\sqrt{3}}Tn^{20}.\label{p:n}
\end{eqnarray}

When solving Eq.~\eqref{ME2}, we truncate the system with $j,l=0,1,\cdots L-1$
and 
\[
p,k=\begin{cases}
2,3,\cdots,K+1, & l=0\\
1,2,\cdots,K, & l=1\\
0,1,\cdots,K-1, & l=2,\cdots,L-1
\end{cases}
\]
to have a system of $N=LK$ moment equations. Enumerating the moment
indices $(l,k)$ as a single index $A=lK+k+\iota=1,2,\cdots,N$, where
\[
\iota=\begin{cases}
-1, & l=0\\
0, & l=1\\
+1, & l=2,\cdots,L-1
\end{cases}
\]
we rewrite Eq.~\eqref{ME2} as 
\begin{equation}
\sum_{B=1}^{N}\Psi_{AB}\frac{\partial n_{B}}{\partial\eta}=\sum_{B=1}^{N}C_{AB}n_{B}+g_{A}.\label{MEll}
\end{equation}
Here the arclength $\ell$ along a magnetic field line is normalized
by the collision length, $d\eta=d\ell/\lambda_{\mathrm{C}}$. The
linear system \eqref{MEll} with constant matrices $\Psi$ and $C$
can be solved by computing the eigensystem of $\Psi^{-1}C$ (see Refs.~\citep{Ji2009HS}
and \citep{Ji2009H1} for details):
\begin{equation}
\sum_{C}(\Psi^{-1}C)_{AB}W_{BC}=k_{D}W_{AC},\label{eig}
\end{equation}
where the eigenvalues $k_{D}$ appear in positive and negative pairs.
The particular solution driven by thermodynamic drives is
\begin{equation}
n_{A}(z)=\sum_{D}\int_{-\infty}^{\infty}K_{AD}(z-z^{\prime})g_{D}(z^{\prime})dz^{\prime},\label{n:K}
\end{equation}
where the kernel functions are defined by
\begin{equation}
K_{AD}(\eta)=\begin{cases}
{\displaystyle -\sum_{\{B|k_{B}>0\}}^{N}\gamma_{AD}^{B}e^{k_{B}\eta}}, & \eta<0,\\
{\displaystyle +\sum_{\{B|k_{B}<0\}}^{N}\gamma_{AD}^{B}e^{k_{B}\eta}}, & \eta>0,
\end{cases}\label{K:e}
\end{equation}
with coefficients 
\begin{equation}
\gamma_{AD}^{B}=\sum_{C}W_{AB}W_{BC}^{-1}\Psi_{CD}^{-1}.\label{a:W}
\end{equation}
For closure moments, we define 
\begin{eqnarray}
\gamma_{hh}^{B} & = & \frac{5}{2}\gamma_{11,11}^{B},\nonumber \\
\gamma_{hR}^{B} & = & -\sqrt{\frac{5}{2}}\sum_{k=1}^{M}a_{\mathrm{ei}}^{1k0}\gamma_{11,1k}^{B}=\gamma_{Rh}^{B},\nonumber \\
\gamma_{h\pi}^{B} & = & {\color{red}-}\sqrt{\frac{5}{3}}\gamma_{11,20}^{B}=\gamma_{\pi h}^{B},\nonumber \\
\gamma_{RR}^{B} & = & \sum_{p,k=1}^{M}a_{\mathrm{ei}}^{10p}a_{\mathrm{ei}}^{1k0}\gamma_{1p,1k}^{B},\nonumber \\
\gamma_{R\pi}^{B} & = & {\color{red}+}\sqrt{\frac{2}{3}}\sum_{k=1}^{M}a_{\mathrm{ei}}^{1k0}\gamma_{20,1k}^{B}=\gamma_{\pi R}^{B},\nonumber \\
\gamma_{\pi\pi}^{B} & = & \frac{4}{3}\gamma_{20,20}^{B},\label{K:K}
\end{eqnarray}
and corresponding $K_{AD}$ by Eq.~\eqref{K:e}. Noting that
\begin{equation}
\gamma_{AD}^{-B}=\begin{cases}
-\gamma_{AD}^{B}, & AD=hh,hR,RR,\pi\pi\equiv\mathrm{even},\\
+\gamma_{AD}^{B}, & AD=h\pi,R\pi\equiv\mathrm{odd},
\end{cases}\label{gam+-}
\end{equation}
where $-B$ denotes the moment index corresponding to $-k_{B}$, we
notice that the kernel functions are even or odd functions:
\begin{equation}
K_{AD}(-\eta)=\begin{cases}
+K_{AD}(\eta), & AD=\mathrm{even}\\
-K_{AD}(\eta), & AD=\mathrm{odd}.
\end{cases}\label{K+-}
\end{equation}
Using the definition of $K_{AD}$ and Eqs.~\eqref{g1k}-\eqref{p:n},
we can write the parallel closures as 
\begin{eqnarray}
h_{\|}(\ell) & = & Tv_{T}\int d\eta^{\prime}\Bigl(-\frac{1}{2}K_{hh}\frac{n}{T}\frac{dT}{d\eta^{\prime}}+K_{hR}Zn\frac{V_{\mathrm{ei}\|}}{v_{T}}-K_{h\pi}\frac{3}{4}n\tau_{\mathrm{ee}}W_{\|}\Bigr),\label{h:}\\
R_{\|}(\ell) & = & -\frac{mn}{\tau_{\mathrm{ei}}}V_{\mathrm{ei}\|}+\frac{mv_{T}}{\tau_{\mathrm{ei}}}\int d\eta^{\prime}\Bigl(-K_{Rh}\frac{n}{2T}\frac{dT}{d\eta^{\prime}}+K_{RR}Zn\frac{V_{\mathrm{ei}\|}}{v_{T}}-K_{R\pi}\frac{3}{4}n\tau_{\mathrm{ee}}W_{\|}\Bigr),\label{R:}\\
\pi_{\|}(\ell) & = & T\int d\eta^{\prime}\Bigl(-K_{\pi h}\frac{n}{T}\frac{dT}{d\eta^{\prime}}+2K_{\pi R}Zn\frac{V_{\mathrm{ei}\|}}{v_{T}}-K_{\pi\pi}\frac{3}{4}n\tau_{\mathrm{ee}}W_{\|}\Bigr).\label{p:}
\end{eqnarray}

The closure calculation from a truncated moment system involves truncation
errors which depend on the collisionality. The inverse collisionality
is often measured by the Knudsen number $k=\lambda_{\mathrm{C}}/|\nabla^{-1}|$.
Since the sinusoidal drives have a constant $k$, we use them to investigate
the truncation errors and convergent behavior of the closures while
increasing the number of moments $N$. Furthermore, in many practical
applications, general drives can be expressed by Fourier series in
a periodic system or its continuum version, Fourier transform, in
a non-periodic system. 

For sinusoidal drives, $T=T_{0}+T_{1}\sin\varphi$, $V_{\|}=V_{0}+V_{1}\sin\varphi$,
and $V_{\mathrm{ei}\|}=V_{\mathrm{ei}}\cos\varphi$, where $\varphi=2\pi\ell/\lambda+\varphi_{0}=k\eta+\varphi_{0}$
and $k=2\pi\lambda_{\mathrm{C}}/\lambda$, assuming that $n$ and
$v_{T}\approx\sqrt{2T_{0}/m}$ are constant and $\nabla\cdot\mathbf{V}_{\perp}\approx0$,
the linearized closures become 
\begin{eqnarray}
h_{\|}(\ell) & = & -\frac{1}{2}nT_{1}v_{T}\hat{h}_{h}\cos\varphi+nT_{0}V_{\mathrm{ei}}\hat{h}_{R}\cos\varphi-nT_{0}V_{1}\hat{h}_{\pi}\sin\varphi,\\
R_{\|}(\ell) & = & -nT_{1}\frac{2\pi}{\lambda}\hat{R}_{h}\cos\varphi-\frac{mnV_{\mathrm{ei}}}{\tau_{\mathrm{ei}}}\hat{R}_{R}\cos\varphi-nmV_{1}\frac{2\pi v_{T}}{\lambda}\hat{R}_{\pi}\sin\varphi,\\
\pi_{\|}(\ell) & = & -nT_{1}\hat{\pi}_{h}\sin\varphi+2nT_{0}\frac{V_{\mathrm{ei}}}{v_{T}}\hat{\pi}_{R}\sin\varphi-nT_{0}\frac{V_{1}}{v_{T}}\hat{\pi}_{\pi}\cos\varphi.
\end{eqnarray}
The dimensionless closures are defined by $\hat{h}_{h}=k\hat{K}_{hh},$
$\hat{h}_{R}=Z\hat{K}_{hR}=\hat{R}_{h},$ $\hat{h}_{\pi}=k\hat{K}_{h\pi}=\hat{\pi}_{h},$
$\hat{R}_{R}=1-Z\hat{K}_{RR},$ $\hat{R}_{\pi}=Z\hat{K}_{R\pi}=\hat{\pi}_{R},$
and $\hat{\pi}_{\pi}=k\hat{K}_{\pi\pi}$, where 
\begin{equation}
\hat{K}_{AD}=\begin{cases}
{\displaystyle \sum_{B=1}^{N}\frac{-\gamma_{AD}^{B}k_{B}}{k_{B}^{2}+k^{2}}}, & AD=\mbox{even}\\
{\displaystyle \sum_{B=1}^{N}\frac{\gamma_{AD}^{B}k}{k_{B}^{2}+k^{2}}}, & AD=\mbox{odd},
\end{cases}\label{K:m}
\end{equation}
which are derived from Eq.~\eqref{K:e}, Eq.~\eqref{gam+-}, and
\begin{equation}
\int K_{AD}(\eta-\eta^{\prime})\cos(k\eta^{\prime}+\varphi_{0})d\eta^{\prime}=\begin{cases}
\hat{K}_{AD}\cos\varphi, & AD=\mbox{even},\\
{\displaystyle \hat{K}}_{AD}\sin\varphi, & AD=\mbox{odd}.
\end{cases}
\end{equation}

\section{Fitted kernel functions for integral closures\label{sec:pll}}

The kernel functions obtained from $N$ moment equations, Eq.~\eqref{K:e},
(i) consist of $N/2$ terms of exponential functions, and (ii) are
inaccurate for $\eta\lesssim\eta_{\mathrm{c}}$ where $\eta_{\mathrm{c}}$
decreases as $N$ increases (e.g. $\eta_{\mathrm{c}}\sim0.01$ for
$N=6400$). The inaccuracy for small $\eta$ introduces an error in
the closure calculation for large wave number $k\gtrsim k_{\mathrm{c}}$.
For example, in the case of the parallel heat flow with $Z=1$ (see
Fig. 2 of Ref.~\citep{Ji2014H1}), the $N=100$ result deviates less
than 1\% from the $N=400$ result for $k\lesssim5$. This means that
the $N=400$ result is accurate within much less than 1\% error for
$k\lesssim5$. Similarly, the $N=400$ result agrees with the $N=1600$
result for $k\lesssim20$ and the $N=1600$ result agrees with the
$N=6400$ result for $k\lesssim80$. As a conservative estimate, $k_{\mathrm{c}}\sim80$
and the $N=6400$ heat flow closure is practically exact for $k\lesssim k_{\mathrm{c}}$.
This convergence scheme can be used to estimate how many parallel
moments are needed for a given $k$ value. To be accurate within 1\%
error, $N=100$ is required for $k\sim5$, $N=400$ for $k\sim20$,
$N=1600$ for $k\sim80$, and so on. Note that the $N=6400$ kernels
consist of 3200 terms and are accurate only for $k\lesssim80$. Therefore,
it is desirable to obtain simple fitted functions that accurately
represent the moment-solution kernels for $\eta\gtrsim\eta_{\mathrm{c}}$,
and the collisionless kernels for $\eta\lesssim\eta_{\mathrm{c}}$.
We obtained the fitted kernels for $Z=1$ in Ref.~\citep{Ji2014H1}
and extend to $Z=2,3,\cdots,10$ in this work. 

In the collisional limit, the parallel closures for arbitrary $Z$
are~\citep{Ji2013H} 
\begin{eqnarray}
h_{\|} & = & -\hat{\kappa}_{\|}\frac{nT\tau_{\mathrm{ee}}}{m}\partial_{\|}T+\hat{\beta}_{\|}nTV_{\mathrm{ei}\|},\label{hc:}\\
R_{\|} & = & -\hat{\beta}_{\|}n\partial_{\|}T-\hat{\alpha}_{\|}\frac{mn}{\tau_{\mathrm{ei}}}V_{\mathrm{ei}\|},\label{Rc:}\\
\pi_{\|} & = & -\hat{\eta}_{0}nT\tau_{\mathrm{ee}}W_{\|}.\label{pc:}
\end{eqnarray}
In the collisionless limit, the closures are determined from the asymptotic
behavior of the kernels for $\eta\ll1$ 
\begin{eqnarray}
K_{hh}(\eta) & \approx & -\frac{18}{5\pi^{3/2}}(\ln|\eta|+\gamma_{h}),\label{hh0}\\
K_{h\pi}(\eta) & \approx & \frac{1}{5},\label{hp0}\\
K_{\pi\pi}(\eta) & \approx & -\frac{4}{5\pi^{1/2}}(\ln|\eta|+\gamma_{\pi}),\label{pp0}
\end{eqnarray}
where $\gamma_{h}$ and $\gamma_{\pi}$ are constants~\citep{Ji2013HJ}.
For the friction related kernels $K_{hR}$, $K_{RR}$, and $K_{R\pi}$,
extrapolating the 6400 moment solution with the constraint Eq.~(\ref{hc:})
will be accurate enough since the corresponding closures vanish as
$\tau\rightarrow\infty$ ($k\rightarrow\infty,$ in the collisionless
limit). 

All kernel functions are fitted to a single function with the same
form of $Z=1$ kernels adopted, 
\begin{equation}
K_{AB}(\eta)=-[d+a\exp(-b\eta^{c})]\ln[1-\alpha\exp(-\beta\eta^{\gamma})].\label{Kfit}
\end{equation}
The parameters \emph{a, b, c, d,} $\alpha$, $\beta$, and $\gamma$
are listed in Table \ref{t:K}. \renewcommand\arraystretch{0.7}

\begin{table}
\begin{tabular}{cccccccccccc}
\hline 
$K_{AB}$ & $Z$ & $1$ & 2 & 3 & 4 & 5 & 6 & 7 & 8 & 9 & 10\tabularnewline
\hline 
\multirow{7}{*}{$K_{hh}$} & $a$ & -3.85 & -3.61 & -4.02 & -4.50 & -5.52 & -6.98  & -9.59  & -14.8  & -24.2 &  -39.0\tabularnewline
 & $b$ & 0.248 & 0.387 & 0.590 & 0.746 & 0.796 & 0.776 & 0.686 & 0.528 & 0.377 & 0.267\tabularnewline
 & $c$ & 0.680 & 0.551 & 0.537 & 0.569 & 0.581 & 0.583 & 0.583 & 0.583 & 0.583 & 0.583\tabularnewline
 & $d$ & 5.40 & 5.47 & 6.07 & 6.66 & 7.74 & 9.28 & 11.9 & 17.1 & 26.5 & 41.4\tabularnewline
 & $\alpha$ & 1 & 1 & 1 & 1 & 1 & 1 & 1 & 1 & 1 & 1\tabularnewline
 & $\beta$ & 2.02 & 2.49 & 2.91 & 3.20 & 3.46 & 3.70 & 3.93 & 4.18 & 4.43 & 4.65\tabularnewline
 & $\gamma$ & 0.417 & 0.348 & 0.316 & 0.300 & 0.291 & 0.281 & 0.279 & 0.277 & 0.276 & 0.275\tabularnewline
\hline 
\multirow{7}{*}{$K_{hR}$} & $a$ & 6.37 & 6.76 & 5.63 & 5.34 & 5.61 & 6.31 & 8.22 & 11.3 & 17.3 & 27.9\tabularnewline
 & $b$ & 5.12 & 5.72 & 6.09 & 6.53 & 6.85 & 7.06 & 7.31 & 7.51 & 7.61 & 7.71\tabularnewline
 & $c$ & 0.160 & 0.179 & 0.219 & 0.240 & 0.239 & 0.227 & 0.205 & 0.181 & 0.154 & 0.126\tabularnewline
 & $d$ & 0.100 & 0.187 & 0.339 & 0.440 & 0.465 & 0.457 & 0.411 & 0.374 & 0.325 & 0.278\tabularnewline
 & $\alpha$ & 1 & 1 & 1 & 1 & 1 & 1 & 1 & 1 & 1 & 1\tabularnewline
 & $\beta$ & 1.00 & 1.73 & 2.50 & 2.96 & 3.19 & 3.33 & 3.37 & 3.39 & 3.37 & 3.34\tabularnewline
 & $\gamma$ & 0.583 & 0.465 & 0.387 & 0.346 & 0.332 & 0.326 & 0.327 & 0.327 & 0.328 & 0.329\tabularnewline
\hline 
\multirow{7}{*}{$K_{h\pi}$} & $a$ & -0.229 & -0.179 & -0.144 & -0.133 & -0.130 & -0.137 & -0.150 & -0.169 & -0.212 & -0.239\tabularnewline
 & $b$ & 2.26 & 3.08 & 3.72 & 4.35 & 4.72 & 4.94 & 5.05 & 5.12 & 5.15 & 5.38\tabularnewline
 & $c$ & 0.594 & 0.596 & 0.594 & 0.588 & 0.569 & 0.562 & 0.556 & 0.551 & 0.548 & 0.543\tabularnewline
 & $d$ & 0.363 & 0.280 & 0.240 & 0.225 & 0.210 & 0.220 & 0.241 & 0.269 & 0.308 & 0.334\tabularnewline
 & $\alpha$ & 0.775 & 0.862 & 0.875 & 0.886 & 0.918 & 0.910 & 0.889 & 0.865 & 0.875 & 0.878\tabularnewline
 & $\beta$ & 1.49 & 1.69 & 1.81 & 1.97 & 2.12 & 2.32 & 2.53 & 2.76 & 3.03 & 3.23\tabularnewline
 & $\gamma$ & 0.478 & 0.460 & 0.454 & 0.442 & 0.432 & 0.415 & 0.399 & 0.380 & 0.362 & 0.351\tabularnewline
\hline 
\multirow{7}{*}{$K_{RR}$} & $a$ & 305 & 322 & 342 & 363 & 386 & 406 & 431 & 450 & 470 & 489\tabularnewline
 & $b$ & 8.30 & 8.67 & 8.90 & 9.09 & 9.23 & 9.32 & 9.40 & 9.49 & 9.52 & 9.54\tabularnewline
 & $c$ & 0.139 & 0.140 & 0.141 & 0.142 & 0.143 & 0.143 & 0.144 & 0.144 & 0.144 & 0.144\tabularnewline
 & $d$ & 0.362 & 0.459 & 0.576 & 0.686 & 0.830 & 0.972 & 1.14 & 1.30 & 1.47 & 1.67\tabularnewline
 & $\alpha$ & 1 & 1 & 1 & 1 & 1 & 1 & 1 & 1 & 1 & 1\tabularnewline
 & $\beta$ & 3.24 & 4.11 & 4.75 & 5.23 & 5.68 & 6.06 & 6.39 & 6.71 & 6.97 & 7.24\tabularnewline
 & $\gamma$ & 0.349 & 0.314 & 0.290 & 0.272 & 0.258 & 0.248 & 0.237 & 0.232 & 0.225 & 0.219\tabularnewline
\hline 
\multirow{7}{*}{$K_{R\pi}$} & $a$ & 0.102 & 0.125 & 0.147 & 0.169 & 0.186 & 0.209 & 0.224 & 0.239 & 0.253 & 0.263\tabularnewline
 & $b$ & 0.528 & 0.724 & 0.898 & 1.06 & 1.22 & 1.30 & 1.51 & 1.61 & 1.77 & 1.91\tabularnewline
 & $c$ & 0.961 & 0.948 & 0.922 & 0.901 & 0.887 & 0.864 & 0.848 & 0.832 & 0.823 & 0.818\tabularnewline
 & $d$ & 0.198 & 0.212 & 0.225 & 0.230 & 0.231 & 0.225 & 0.220 & 0.213 & 0.207 & 0.202\tabularnewline
 & $\alpha$ & 1 & 1 & 1 & 1 & 1 & 1 & 1 & 1 & 1 & 1\tabularnewline
 & $\beta$ & 2.45 & 3.06 & 3.52 & 3.87 & 4.15 & 4.38 & 4.57 & 4.73 & 4.88 & 5.02\tabularnewline
 & $\gamma$ & 0.408 & 0.370 & 0.347 & 0.332 & 0.322 & 0.313 & 0.307 & 0.303 & 0.299 & 0.294\tabularnewline
\hline 
\multirow{7}{*}{$K_{\pi\pi}$} & $a$ & 0.470 & 0.598 & 0.700 & 0.762 & 0.804 & 0.839 & 0.857 & 0.873 & 0.878 & 0.883\tabularnewline
 & $b$ & 1.06 & 1.19 & 1.31 & 1.45 & 1.59 & 1.72 & 1.85 & 1.97 & 2.08 & 2.18\tabularnewline
 & $c$ & 0.661 & 0.607 & 0.580 & 0.566 & 0.557 & 0.551 & 0.546 & 0.543 & 0.541 & 0.539\tabularnewline
 & $d$ & 0.357 & 0.275 & 0.207 & 0.166 & 0.139 & 0.118 & 0.106 & 0.096 & 0.091 & 0.087\tabularnewline
 & $\alpha$ & 1 & 1 & 1 & 1 & 1 & 1 & 1 & 1 & 1 & 1\tabularnewline
 & $\beta$ & 1.66 & 1.97 & 2.17 & 2.34 & 2.49 & 2.61 & 2.74 & 2.85 & 2.97 & 3.08\tabularnewline
 & $\gamma$ & 0.546 & 0.517 & 0.498 & 0.487 & 0.479 & 0.472 & 0.469 & 0.466 & 0.465 & 0.465\tabularnewline
\hline 
\end{tabular}\caption{Fitted parameters in Eq.~\eqref{Kfit} for $Z=1,2,\cdots,10$.}
\label{t:K}
\end{table}

In computing the fitted kernel parameters there are many least-squares
local minima which accurately represent the convergent kernels ($\eta\gtrsim0.01)$.
We use sinusoidal drives to assess the accuracy of fitted kernels.
The closures computed from fitted kernels are compared with 6400 moment
closures in the convergent regime ($k\lesssim80$). Note that the
fitted parameters automatically satisfy kernels for $\eta\lesssim0.01$
forced by Eqs.~\eqref{hh0}-\eqref{pp0} and therefore closures for
$K_{hh},\;K_{h\pi},$ and $K_{\pi\pi}$ are accurate in the collisionless
limit. For friction related kernels $K_{hR},\;K_{R\pi},$ and $K_{RR}$,
the closures are ignorable in the collisionless (no friction) limit. 

In the interest of including noninteger effective ion charge numbers,
we choose sets of least-squres fitting parameters that change smoothly
in $Z$. Although some parameters for $Z=1$ in this work are different
from the ones in Ref.~\citep{Ji2014H1}, they provide similar accuracy
for closure calculations. For a noninteger ion-charge number $Z_{\mathrm{eff}}$,
$Z<Z_{\mathrm{eff}}<Z+1$, a simple linear interpolation of parameters
$A=a,b,c,d,\beta,\gamma$ 
\begin{equation}
A_{Z\mathrm{eff}}=(1+Z-Z_{\mathrm{eff}})A_{Z}+(Z_{\mathrm{eff}}-Z)A_{Z+1}\label{intpl}
\end{equation}
results in accurate results. We note that using the constraints \eqref{hh0}-\eqref{pp0}
instead of interpolating all parameters results in higher accuracy.
We obtain $a$ from other interpolated parameters for $K_{hh}$ and
$K_{\pi\pi}$ 
\begin{eqnarray}
a & = & \frac{18}{5\pi^{3/2}\gamma}-d\mbox{ for }K_{hh},\label{ahh}\\
a & = & \frac{4}{5\pi^{1/2}\gamma}-d\mbox{ for }K_{\pi\pi},\label{app}
\end{eqnarray}
and $\alpha$ for $K_{h\pi}$
\begin{equation}
\alpha=1-\exp\frac{-1}{5(a+d)}\mbox{ for }K_{h\pi}.\label{ahp}
\end{equation}
\begin{table}
\begin{tabular}{>{\centering}p{0.12\textwidth}>{\centering}p{0.12\textwidth}>{\centering}p{0.12\textwidth}>{\centering}p{0.12\textwidth}>{\centering}p{0.12\textwidth}>{\centering}p{0.12\textwidth}>{\centering}p{0.12\textwidth}}
\hline 
$Z$ & $\hat{K}_{hh}$ & $\hat{K}_{hR}$ & $\hat{K}_{h\pi}$ & $\hat{K}_{RR}$ & $\hat{K}_{R\pi}$ & $\hat{K}_{\pi\pi}$\tabularnewline
\hline 
1 & 1.0 & 0.6 & 0.6 & 0.7 & 1.0 & 0.5\tabularnewline
1.5 & 2.4 & 3.2 & 2.7 & 4.0 & 3.2 & 1.6\tabularnewline
2 & 2.8 & 0.9 & 1.0 & 0.7 & 0.9 & 0.8\tabularnewline
2.5 & 3.0 & 4.4 & 1.8 & 2.4 & 1.5 & 1.1\tabularnewline
3 & 4.9 & 1.9 & 0.7 & 0.7 & 0.6 & 0.6\tabularnewline
3.5 & 4.3 & 2.3 & 1.1 & 1.5 & 0.9 & 1.0\tabularnewline
4 & 4.8 & 4.3 & 0.8 & 0.3 & 0.3 & 0.4\tabularnewline
4.5 & 4.4 & 4.1 & 1.2 & 0.8 & 0.9 & 0.5\tabularnewline
5 & 4.7 & 4.4 & 0.8 & 0.2 & 0.5 & 0.4\tabularnewline
5.5 & 4.2 & 3.7 & 0.8 & 0.3 & 0.8 & 0.4\tabularnewline
6 & 4.6 & 3.9 & 0.8 & 0.5 & 0.5 & 0.4\tabularnewline
6.5 & 3.1 & 1.0 & 0.8 & 0.3 & 0.6 & 0.2\tabularnewline
7 & 3.1 & 0.8 & 1.0 & 0.4 & 0.7 & 0.5\tabularnewline
7.5 & 2.8 & 1.5 & 1.4 & 0.2 & 0.5 & 0.4\tabularnewline
8 & 3.0 & 0.9 & 2.0 & 0.2 & 0.7 & 0.4\tabularnewline
8.5 & 3.7 & 4.0 & 2.3 & 0.3 & 0.9 & 0.4\tabularnewline
9 & 3.4 & 1.8 & 2.5 & 0.2 & 0.9 & 0.5\tabularnewline
9.5 & 3.4 & 3.1 & 2.6 & 0.3 & 0.7 & 0.3\tabularnewline
10 & 3.4 & 3.2 & 2.7 & 0.8 & 0.9 & 0.3\tabularnewline
\hline 
\end{tabular}\caption{Maximum percentage deviation of closures computed with fitted kernels
from 6400 moment closures in the convergent regime $\eta\lesssim80$
for $1\le Z\le10$. For half integers, kernel parameters are computed
by linear interpolation.}
\label{t:per}
\end{table}
The maximum deviations from the closures in the convergent regime
$(k\lesssim80)$ are shown for integers and half-integers in Table
\ref{t:per}. The maximum deviations usually occur at $k$ where the
closure values are close to zero. For a noninteger $Z<Z_{\mathrm{eff}}<Z+1$,
the error is less than the maximum of errors at $Z$, $Z+1/2$, and
$Z+1$. The maximum errors are less than 5\% at the worst case for
any arbitrary $1\le Z\le10$. 

\begin{figure}
\includegraphics{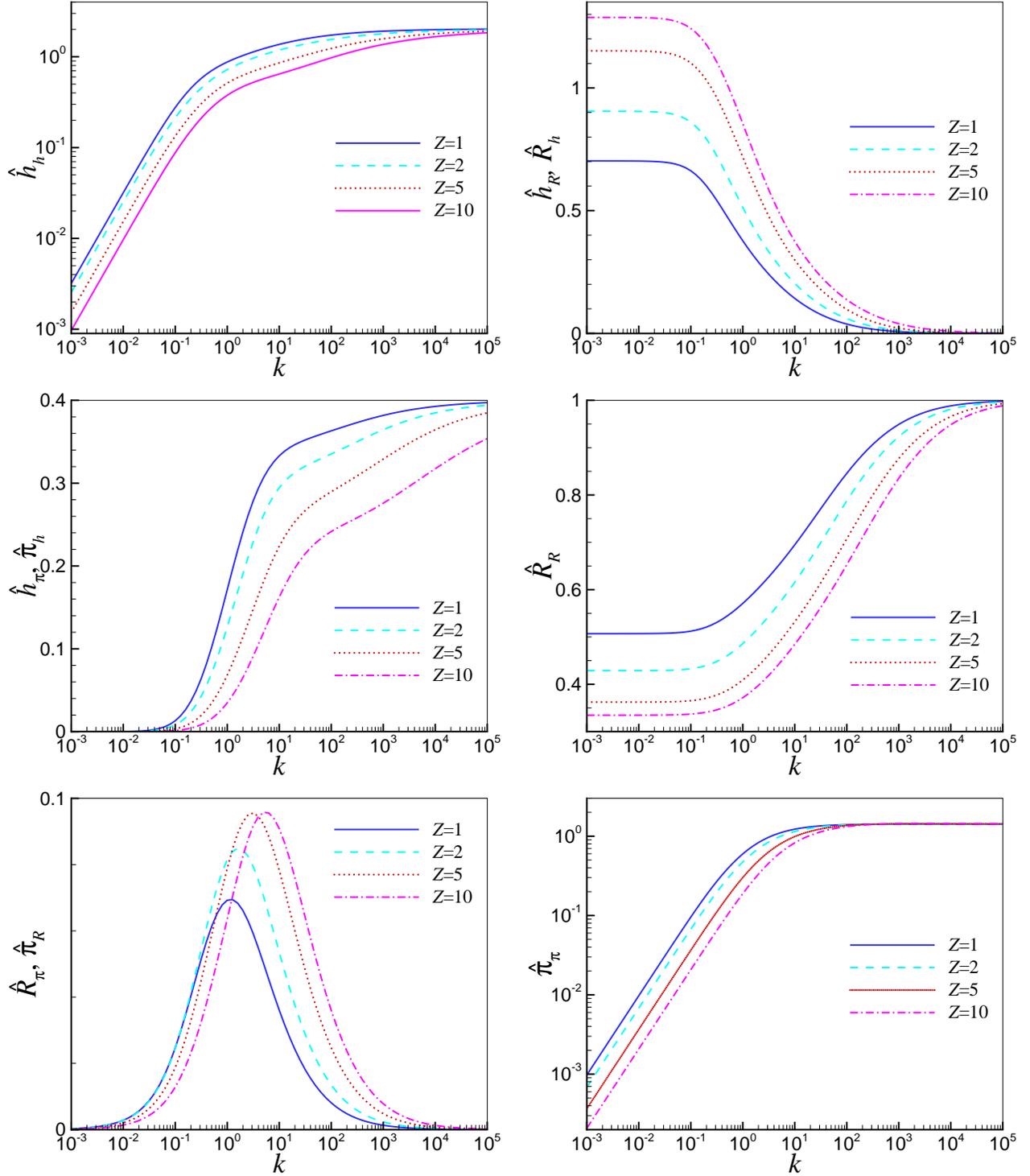}\caption{(Color online) Closures for sinusoidal drives computed from fitted
kernels for $Z=1,2,5,$ and 10.}
\label{fig:n}
\end{figure}
Fig.~\ref{fig:n} shows typical behavior of closures due to sinusoidal
drives for various $Z$. In the collisional $(k\ll1)$ limit, the
closures approach the corresponding high-collisionality values for
each $Z$~\citep{Ji2013H}. In the collisionless ($k\rightarrow\infty)$
limit, the closures approach $Z$-independent collisionless-limit
values~\citep{Ji2013HJ}. Although the maximum errors are verified
to be less than 5\% for $k\lesssim80$, the errors may be larger than
5\% for $k\gtrsim80$. Since the exact values are unknown in this
regime (the 6400 moment closures do not converge) we can only estimate
the accuracy of closures from the shape of curves. In this regime,
the change of closure values $\hat{h}_{h}$ for $Z=10$ and $h_{\hat{\pi}}\;(\hat{\pi}_{h})$
for $Z=5,10$ seems slightly eccentric. Nevertheless, the errors are
expected to be not much greater than 5\%, since the closure values
eventually approach the theoretical values in the collisionless limit.

\section{Summary\label{sec:Dis}}

In obtaining simple fitted kernels for electron parallel closures,
we extended the $Z=1$ calculation to $Z=2,\cdots10$. Since parameters
change smoothly in $Z$, linear interpolation of parameters at $Z$
and $Z+1$ yields the parameter for noninteger $Z<Z_{\mathrm{eff}}<Z+1$
with the same order of accuracy in computing closures.

The same method can be applied to ion parallel closures. As shown
in Refs.~\citep{Ji2009H1,Ji2015H}, inclusion of the ion-electron
collision operator is necessary. The ion-electron operator introduces
two independent parameters, the mass ratio combined with the ion charge
number and the temperature ratio. Fitted kernels for ion parallel
closures will appear in future work. 

\section*{Acknowledgments}

One of the authors (Ji) would like to thank the Fusion and Plasma
Application Laboratory (FUSMA) Team at Seoul National University for
their kind support during his visit. The research was supported by
the U.S. DOE under grant nos. DE-SC0014033, DE-FG02-04ER54746, DE-FC02-04ER54798,
and DE-FC02-05ER54812, and by the National R\&D Program through the
National Research Foundation of Korea (NRF), funded by the Ministry
of Science, ICT \& Future Planning (No. 2014-M1A7A1A03045368), and
by the project PE15090 of Korea Polar Research Institute. This work
was performed in conjunction with the Plasma Science and Innovation
(PSI) Center and the Center for Extended MHD Modeling (CEMM).

\end{document}